\documentclass[showpacs,aps,twocolumn]{revtex4}
\usepackage{epsf}
\newcommand{\mb}[1]{ { \mbox{\boldmath{$#1$}}}  } 
 
\begin{document}
    
\title{Enhancements of Andreev conductance induced by the photon/vibron scattering}

\author{J.\ Bara\'nski$^{1,2}$  and  T.\ Doma\'nski$^{1}$}
\affiliation{
       $^{1}$Institute of Physics, M.\ Curie-Sk\l odowska University, 
       20-031 Lublin, Poland \\
       $^{2}$Institute of Physics, Polish Academy of Sciences, 
       02-668 Warsaw, Poland}

\date{\today}

\begin{abstract}
We analyze the subgap spectrum and transport properties of the quantum 
dot embedded between one superconducting and another metallic reservoirs 
and additionally coupled to an external boson mode. Emission/absorption  
of the bosonic quanta induces a series of the subgap Andreev states,
that eventually interfere with each other. We discuss their signatures 
in the differential conductance both, for the linear and nonlinear regimes.
\end{abstract}

\maketitle

\section{Introduction}
The bosonic modes, like  photons \cite{Platero_Aguado_2004} 
or  vibrational degrees of freedom \cite{Galperin_2007}, can  
strongly affect electron tunneling through the nanoscopic 
systems \cite{Koch_2005}. When a level spacing of nanoobject 
is large in comparison to the boson energy $\omega_{0}$ 
and a line-broadening is sufficiently narrow, a series of 
the side-peaks \cite{Fransson_book} may appear due 
to emission/absorption of the bosonic quanta. Such features 
(spaced by  $\omega_{0}$) have been really observed in measurements 
of the differential conductance for several nanojunctions 
\cite{Sapmaz_2006,Leturcq_2009,Beebe_2006,Pasupathy_2005}.
 
Similar bosonic modes are currently studied also in the systems, where 
the quantum dots/impurities are coupled with superconducting reservoirs 
\cite{Cho_1999,Song_2008,Zazunov_2006,Fransson_2010,Timm_2012,Zhang_2009,
Bai_2011,Sun_2012,Zitko_2012,Rudzinski_2014,Komnik_13,Wang_2013}. 
Since the proximity effect spreads electron pairing onto these quantum 
dots, the bosonic features manifest themselves in a quite peculiar way. 
They could be observed by  the Josephson 
\cite{Zazunov_2006,Fransson_2010,Timm_2012} and the Andreev spectroscopies 
\cite{Cho_1999,Zhang_2009,Bai_2011,Sun_2012,Zitko_2012,Rudzinski_2014},
in photon-assisted subgap tunneling \cite{Song_2008}, 
transient phenomena  \cite{Komnik_13},
or in prototypes of the nano-refrigerators operating due to 
the multi-phonon Andreev scattering \cite{Wang_2013}.

First of all, in a subgap regime (assuming $\omega_{0}$ smaller 
than energy gap $\Delta$ of superconductor) the bosonic features 
are expected to be more numerous than in the normal state. 
This is a consequence the proximity effect, mixing the particle 
and hole excitations. Secondly, it has been shown numerically 
\cite{Cho_1999,Sun_2012,Wang_2013} that the linear (zero-bias) 
Andreev conductance exhibits the bosonic features spaced by 
a half of $\omega_{0}$. To our knowledge, this intriguing 
theoretical result was neither clarified on physical arguments 
nor checked experimentally. Verification would be 
feasible by the tunneling spectroscopy using e.g.\ 
low-frequency vibrations of some heavy molecules
or slowly-varying  {\em ac} electromagnetic field. 
Let us emphasize, that such low-energy boson mode need 
not be related with any pairing mechanism of the
superconducting reservoir. 

The purpose of our paper is to provide a simple analytical 
argument, explaining the reduced frequency $\omega_{0}/2$
of the bosonic features in the linear Andreev conductance 
versus the gate-voltage. We also study in detail the multiple 
subgap states originating from the boson emission/absorption 
processes. We analyze their signatures both in the quantum 
dot spectrum and the tunneling transmission. The latter quantity
can be probed by the (low-temperature) differential conductance 
as a function of the source-drain bias. We predict that the 
multiple Andreev states could be seen with a period, 
dependent on the gate voltage.

\begin{figure}
\epsfxsize=9cm\centerline{\epsffile{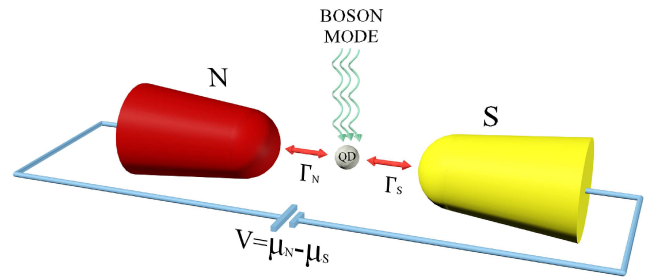}}
\caption{(color online) A scheme of the quantum dot between 
the metallic (N) and  superconducting (S) electrodes and 
coupled to the monochromatic boson (phonon or photon) mode.}
\label{scheme}
\end{figure}

For calculations we consider the setup displayed in figure 
\ref{scheme}. It can be practically realized in a single 
electron transistor (SET) using e.g.\ the carbon nanotube 
suspended between the external electrodes (like in Refs 
\cite{Sapmaz_2006,Leturcq_2009}). Another possibility 
could be the scanning tunneling microscope (STM), where 
the conducting tip (N) probes some vibrating quantum 
impurity (QD) hosted in a superconducting (S) substrate 
\cite{Balatsky-06}. In both SET and STM configurations 
such boson mode can be eventually related to external 
{\em ac} field.

In what follows we introduce the Hamiltonian and discuss the method 
for treating the bosonic mode. We next investigate the bosonic 
signatures in the QD spectrum and in the subgap Andreev conductance. 
For clarity, we focus on the limit $\Gamma_{N}\ll \omega_{0}$ 
whereas the second coupling $\Gamma_{S}$ can be arbitrary. 
In the last section we address the correlation effects.

\section{Microscopic model}

For microscopic description of  the tunneling scheme shown 
in Fig.\ \ref{scheme} we use the Anderson impurity model
\begin{eqnarray} 
\hat{H} =   \hat{H}_{N}+\hat{H}_{S} + \hat{H}_{mol} + \hat{H}_{T} .
\label{model} 
\end{eqnarray}
$\hat{H}_{N(S)}$ refers to the normal (superconducting) lead, 
$\hat{H}_{mol}$ describes the molecular quantum dot (i.e.\ the 
localized electrons coupled with the boson mode) and $\hat{H}_{T}$ 
is a hybridization between the QD and itinerant electrons. 
We  treat the  normal electrode as a free Fermi gas 
$\hat{H}_{N}=\sum_{{\bf k},\sigma} \xi_{{\bf k}N} \hat{c}_{{\bf k} 
\sigma N}^{\dagger} \hat{c}_{{\bf k}\sigma N}$ and describe 
the other superconducting lead by the BCS Hamiltonian $\hat{H}_{S} 
\!=\!\sum_{{\bf k},\sigma}  \xi_{{\bf k}S} 
\hat{c}_{{\bf k} \sigma S }^{\dagger}  \hat{c}_{{\bf k} \sigma S} 
\!-\! \Delta \sum_{\bf k} ( \hat{c}_{{\bf k} \uparrow S }^{\dagger} 
\hat{c}_{-{\bf k} \downarrow S }^{\dagger} \!+\! \hat{c}_{-{\bf k} 
\downarrow S} \hat{c}_{{\bf k} \uparrow S})$. The annihilation 
(creation) operators $\hat{c}_{{\bf k} \sigma \beta} ^{({\dagger})}$ 
correspond to mobile  $\beta=N,S$ electrons with spin $\sigma=\uparrow,
\downarrow$ and energy $\xi_{{\bf k}\beta}\!=\!\varepsilon_{{\bf k}
\beta} \!-\!\mu_{\beta}$ measured with respect to the chemical 
potential $\mu_{\beta}$. Nonequlibrium conditions can be driven 
by the bias $V=\mu_{L}-\mu_{R}$ and/or temperature difference 
$T_{L}\neq T_{R}$. The induced currents depend qualitatively 
on the hybridization 
$\hat{H}_{T} =  \sum_{{\bf k},\sigma,\beta} 
\left( V_{{\bf k} \beta} \; \hat{d}_{\sigma}^{\dagger}  
\hat{c}_{{\bf k} \sigma \beta } + \mbox{\rm H.c.} \right)$ 
and on parameters of the molecular quantum dot 
\begin{eqnarray} 
\hat{H}_{mol} = \varepsilon \sum_{\sigma} \hat{n}^{\dagger}
_{d\sigma} + U  \hat{n}_{d \uparrow} \hat{n}_{d \downarrow}
+ \omega_0 \hat{a}^{\dagger}\hat{a}+\lambda \sum_{\sigma}
\hat{n}_{d \sigma}(\hat{a}^{\dagger}\!+\hat{a}) .
\label{QD} 
\nonumber
\end{eqnarray}  
The number operator $\hat{n}_{d\sigma} = \hat{d}^{\dagger}
_{\sigma} \hat{d}_{\sigma}$ counts the localized electrons with spin 
$\sigma$, $\varepsilon$ is the QD energy level and $U$ denotes 
the Coulomb potential between opposite spin electrons. The boson 
field (described by $\hat{a}^{(\dagger)}$ operators) is assumed
as a monochromatic mode $\omega_{0}$ and its coupling 
with the QD electrons is denoted by $\lambda$.

\section{Multiple subgap states}

There are three main obstacles in determining the effective
energy spectrum and the tunneling transmission of our system:
i) the electron-boson coupling $\lambda$, ii) the proximity
induced on-dot pairing (due to $\Delta$), and iii) the correlation
effects caused by the Coulomb repulsion $U$. The most reliable way 
for studying them on equal footing would be possible within 
the numerical renormalization group \cite{Zitko_2012} approach, 
however such method encounters problems in estimating the Andreev 
transmission. To get some insight into the spectrum and transport 
properties we start by neglecting the correlations and then 
(in the last section) treat them using the superconducting 
atomic limit solution. 

Following \cite{Cho_1999,Song_2008,Zazunov_2006,Fransson_2010,Timm_2012,
Zhang_2009,Bai_2011,Sun_2012,Zitko_2012,Rudzinski_2014,Wang_2013}
we apply the unitary transformation $e^{\hat{S}} \hat{H} 
e^{-\hat{S}}=\hat{\tilde{H}}$ to decouple the electron 
from boson quasiparticles. With the Lang-Firsov generating 
operator \cite{Lang_Firsov}
\begin{eqnarray}
\hat{S}=\frac{\lambda}{\omega_0} \sum_{\sigma}\hat{n}_{d\sigma}
 \left( \hat{a}^{\dagger}-\hat{a} \right)
\label{Lang_S}
\end{eqnarray}
the molecular Hamiltonian (\ref{QD}) is transformed to
\begin{eqnarray}
\hat{\tilde{H}}_{mol} = \sum_{\sigma} \tilde{\varepsilon} 
\hat{\tilde{d}}_{\sigma}^{\dagger} \hat{\tilde{d}}_{\sigma}
+\tilde{U} \hat{\tilde{n}}_{\downarrow} \hat{\tilde{n}}_{\uparrow} 
+\omega_{0} \hat{a}^{\dagger} \hat{a} ,
\label{afterLF}
\end{eqnarray}
where the energy level is lowered by the polaronic shift 
$\tilde{\epsilon}=\varepsilon\!-\!\lambda^2 / \omega_0$ and 
the effective potential $\tilde{U}\!=\!U-2\lambda^2 / \omega_0$. 
Boson operators are shifted $\hat{\tilde{a}}^{(\dagger)}=
\hat{a}^{(\dagger)}-\frac{\lambda}{\omega_0} \sum_{\sigma} 
\hat{d}_{\sigma}^{\dagger} \hat{d}_{\sigma}$ whereas fermions 
are {\em dressed} with the polaronic cloud 
\begin{eqnarray}
\hat{\tilde{d}}_{\sigma}^{(\dagger)}=\hat{d}_{\sigma}^{(\dagger)} 
\hat{X}^{(\dagger)} ,\hspace{1cm}  
\hat{X}=e^{-(\lambda / \omega_0)(\hat{a}^{\dagger}-\hat{a})} .
\label{X}
\end{eqnarray}
Reservoirs $\hat{H}_{\beta}$ are invariant on the unitary 
transformation (\ref{Lang_S}) but the operator $\hat{X}$ appears 
in the hybridization term $\hat{\tilde{H}}_{T}$. For simplicity 
we absorb it into the effective coupling constants $\Gamma_{\beta}
=2\pi \sum_{\bf k} |V_{{\bf k}\beta}|^2 \; \langle \hat{X}
^{\dagger}\hat{X}\rangle \delta(\omega \!-\! \xi_{{\bf k}\beta})$ 
which can be defined for the wide band limit. 
 
The effective single particle excitation spectrum is given by  
the Green's function 
\begin{eqnarray}
\mb{G}_{\sigma}(\tau_{1},\tau_{2}) = -i \; \left\langle\hat{T}_{\tau} 
\hat{d}_{\sigma}(\tau_{1})\hat{d}_{\sigma}^{\dagger}(\tau_{2}) 
\right\rangle_{\hat{H}} ,
\label{Gmol}
\end{eqnarray}
where $\hat{T}_{\tau}$ denotes the time ordering operator. Since trace 
is invariant on the unitary transformations $\langle ... \rangle
_{\hat{H}} = \langle ... \rangle_{\hat{\tilde{H}}}$ it is convenient
to compute the statistical averages with respect to $\hat{\tilde{H}}$. 
In particular,  (\ref{Gmol}) can be expressed as
\begin{eqnarray}
\mb{G}_{\sigma}(\tau_{1},\tau_{2}) = -i \left\langle\hat{T}_{\tau}
\hat{d}_{\sigma}(\tau_{1})\hat{d}_{\sigma}^{\dagger}(\tau_{2})
\right\rangle_{\hat{\tilde{H}}_{fer}} \!\!\left\langle\hat{T}
_{\tau}\hat{X}(\tau_{1}) \hat{X}^{\dagger}(\tau_{2})\right
\rangle_{\hat{\tilde{H}}_{bos}} 
\label{product}
\end{eqnarray}
because the fermionic and bosonic degrees of freedom are separated
by the Lang-Firsov transformation. From a standard procedure 
\cite{Fransson_book,Mahan_book} one obtains 
\begin{eqnarray}
&&\left\langle \hat{T}_{\tau} \hat{X}(\tau_{1})\hat{X}(\tau_{2})^{\dagger}
\right\rangle_{\hat{\tilde{H}}_{bos}} = \mbox{\rm exp}\left\lbrace 
-(\lambda/\omega_0)^2 \right. \; \times \label{eq_19} \\ 
& & \left. [(1-e^{-i\omega_0 (\tau_{1}-\tau_{2})}) (1+N_p) 
+ (1-e^{i\omega_0 (\tau_{1}-\tau_{2})})N_p]\right\rbrace 
\nonumber
\end{eqnarray}
with the Bose-Einstein distribution $N_p=\left[ e^{\beta \omega_0}
-1\right]^{-1}$. Fourier transform of the Green's function 
(\ref{eq_19}) is found as
\begin{eqnarray}
\mb{G}_{\sigma}(\omega)&=&\sum_{l} \mb{g}_{\sigma}(\omega\!-\!l\omega_{0}) \; 
\; e^{-(\lambda \sqrt{1+2N_p}/\omega_0)^2} 
\label{eq_16} \\ && \times 
e^{l \beta \omega_0/2} I_l \left[ 2(\frac{\lambda}{\omega_0})^{2} 
\sqrt{N_p(1+N_p)} \right]  ,
\nonumber
\end{eqnarray}
where $I_{l}$ denote the modified Bessel functions and 
$\mb{g}_{\sigma}(\tau_{1},\tau_{2}) = -i \left\langle\hat{T}_{\tau}
\hat{d}_{\sigma}(\tau_{1})\hat{d}_{\sigma}^{\dagger}(\tau_{2})
\right\rangle_{\hat{\tilde{H}}_{fer}}$ is the fermionic part 
of (\ref{product}). In the ground state (\ref{eq_16}) simplifies to 
\begin{eqnarray}
\lim_{T \rightarrow 0} \mb{G}_{\sigma}(\omega) = \sum_{l} \mb{g}
_{\sigma}(\omega\!-\!l\omega_{0}) \;\; e^{-g} \; \frac{g^{l}}{l!} 
\label{eq_21} 
\end{eqnarray}
with the adiabatic parameter  $g=(\lambda/\omega_0)^2$. 

Due to the proximity induced on-dot pairing the single particle 
Green's function $\mb{G}_{\uparrow}(\tau_{1},\tau_{2})$ is 
mixed with the (anomalous) propagator
\begin{eqnarray}
&& \mb{F}(\tau_{1},\tau_{2}) = -i
\left\langle\hat{T}_{\tau} \hat{d}_{\downarrow}^{\dagger}(\tau_{1})
\hat{d}_{\uparrow}^{\dagger}(\tau_{2}) \right\rangle_{\hat{H}} =
\label{G_off} \\
&& -i \left\langle\hat{T}_{\tau} \hat{d}_{\downarrow}^{\dagger}
(\tau_{1})\hat{d}_{\uparrow}^{\dagger}(\tau_{2}) \right\rangle_{\hat{\tilde{H}}_{fer}} 
\left\langle\hat{T}_{\tau}\hat{X}^{\dagger}(\tau_{1})
\hat{X}^{\dagger}(\tau_{2})\right\rangle_{\hat{\tilde{H}}_{bos}} .
\nonumber
\end{eqnarray}
This important fact has been remarked in the previous considerations 
of {\em dc} Josephson current \cite{Timm_2012} and it also plays 
significant role for the Andreev spectroscopy (see the next section). 
The boson part of the anomalous propagator (\ref{G_off}) 
takes the following form
\begin{eqnarray}
&&\left\langle \hat{T}_{\tau} \hat{X}^{\dagger}(\tau_{1})
\hat{X}(\tau_{2})^{\dagger} \right\rangle_{bos} 
= \mbox{\rm exp}\left\lbrace -(\lambda/\omega_0)^2 \right. 
\; \times  \\ 
& & \left. [(1+e^{-i\omega_0 (\tau_{1}-\tau_{2})}) (1+N_p) 
+ (1+e^{i\omega_0 (\tau_{1}-\tau_{2})})N_p]\right\rbrace .
\nonumber
\end{eqnarray}
At zero temperature its Fourier transform simplifies  to 
\begin{eqnarray}
\lim_{T \rightarrow 0} \mb{F}(\omega) = \sum_{l} \mb{f}(\omega\!-\!
l\omega_{0}) \;\; e^{-g} \; \frac{(-g)^{l}}{l!}  .
\label{F_func0}
\end{eqnarray}
As regards the fermion part $\mb{f}(\tau_{1},\tau_{2}) = -i \langle
\hat{T}_{\tau}\hat{d}^{\dagger}_{\downarrow}(\tau_{1})\hat{d}
_{\uparrow}^{\dagger}(\tau_{2})\rangle_{\hat{\tilde{H}}_{fer}}$ 
it couples to the Green's function $\mb{g}(\tau_{1},
\tau_{2})$. Their Fourier components obey the Dyson 
equation
\begin{eqnarray} 
&& \left[ \begin{array}{cc}  
\mb{g}(\omega) &  \mb{f}(\omega) \\ \mb{f}^{\star}(-\omega) 
&  -\mb{g}^{\star}(-\omega) \end{array} \right]^{-1}
\label{gf_matrix} \\
&=&\left[ \begin{array}{cc}  \omega\!-\!\tilde{\varepsilon} &  0 \\ 0 &  
\omega\!+\!\tilde{\varepsilon}\end{array}\right] 
- {\mb \Sigma}_{QD}^{0}(\omega) - {\mb \Sigma}_{QD}^{corr}(\omega) ,  
\nonumber 
\end{eqnarray} 
where ${\mb \Sigma}_{d}^{0}$ is the selfenergy matrix of uncorrelated
molecular dot and the second contribution ${\mb  \Sigma}_{d}^{corr}$ 
is due to the effective Coulomb interaction $\tilde{U}$. In the 
wide-band limit the selfenergy ${\mb \Sigma}_{QD}^{0}(\omega)$ 
can be expressed as 
\begin{eqnarray}
\mb{\Sigma}_{QD}^{0}(\omega) =  -i \frac{\Gamma_{N}}{2} \; 
\left( \begin{array}{cc}  
1 & 0 \\ 0 & 1 \end{array} \right)
- \frac{\Gamma_{S}}{2} \gamma(\omega)
\left( \begin{array}{cc}  
1 & \frac{\Delta}{\omega} \\ 
 \frac{\Delta}{\omega}  & 1 
\end{array} \right)
\label{selfenergy_0}
\end{eqnarray} 
with 
\begin{eqnarray}
\gamma(\omega) = \left\{
\begin{array}{ll} 
\frac{\omega}{\sqrt{\Delta^{2}-\omega^{2}}}
& \mbox{\rm for }  |\omega| < \Delta , \\
\frac{i\;|\omega|}{\sqrt{\omega^{2}-\Delta^{2}}}
& \mbox{\rm for }  |\omega| > \Delta .
\end{array} \right.
\label{gamma}
\end{eqnarray} 

\begin{figure}
\epsfxsize=8.5cm\centerline{\epsffile{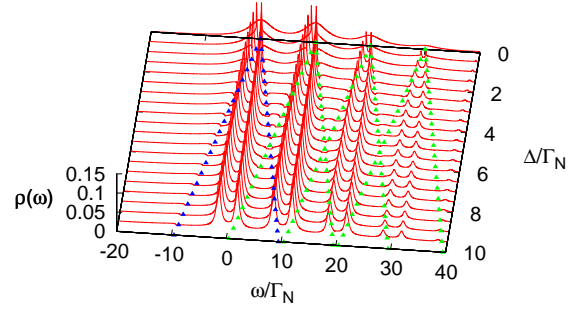}}
\caption{(color online) Energy spectrum $\rho(\omega)$ of the 
uncorrelated quantum dot ($\tilde{U}\!=\!0)$ obtained at $T=0$ 
for $\tilde{\varepsilon}=0$, $g=1$, $\omega_0=10\Gamma_N$. 
The filled triangles at $l\omega_{0}\pm\Delta$ are only
guide to eye. For increasing $\Delta$ the boson peaks split 
into the lower and upper states and their broadening shrinks 
to $\Gamma_{N}$.} 
\label{Fig2}
\end{figure}

\begin{figure}
\epsfxsize=8.5cm\centerline{\epsffile{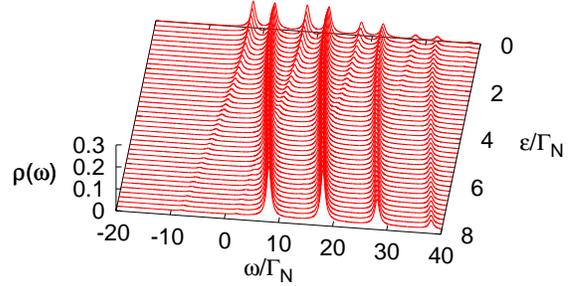}}
\caption{(color online) Spectrum of the uncorrelated quantum 
dot for $\omega_{0}=10 \Gamma_{N}$, $g=1$, $\Gamma_S=4
\Gamma_N$, $T=0$. The neighboring  boson peaks are crossing
at $\omega=(\frac{1}{2}+l)\omega_{0}$
for $\tilde{\varepsilon} \simeq \omega_{0}/2$.}
\label{Fig3}
\end{figure}

\begin{figure}
\epsfxsize=8.5cm\centerline{\epsffile{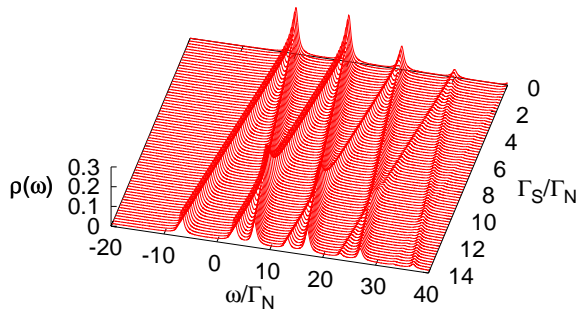}}
\caption{(color online) Spectrum of the uncorrelated quantum dot 
for $\tilde{\varepsilon}=0$, $g=1$, $\omega_0=10\Gamma_N$, $T=0$, 
$\Delta\gg\Gamma_S$. The bosonic features cross each other
at $\omega=(\frac{1}{2}+l)\omega_{0}$ for $\Gamma_{S}=
\omega_{0}$.} 
\label{Fig4}
\end{figure}

We investigated the effective spectral function $\rho(\omega)=-\pi^{-1}
\mbox{\rm Im}\mb{G}(\omega+i0^{+})$ at zero temperature, focusing 
on the intermediate electron-boson coupling $g\sim 1$. Figures 
\ref{Fig2}--\ref{Fig4} show the QD spectrum for $\tilde{U}\!=\!0$, 
neglecting the correlation effects ${\mb  \Sigma}_{d}^{corr}$. 
Influence of the Coulomb potential $\tilde{U}$ is discussed in section V.

Fig \ref{Fig2} illustrates evolution of the bosonic features with 
respect to the superconductor gap $\Delta$. In the normal state 
(for $\Delta=0$) such lorentzian peaks are located at $\omega=\tilde
{\varepsilon}+l\omega_{0}$ (with integer $l\geq0$) and their
broadening is $\Gamma_{N}+\Gamma_{S}$.  For finite $\Delta$ all 
peaks split into the lower and upper ones due to the induced 
on-dot pairing. In the extreme limit $\Delta\gg\Gamma_{S}$ 
the selfenergy $\mb{\Sigma}_{QD}^{0}(\omega)$ becomes static
\begin{eqnarray}
\lim_{\Delta \gg \Gamma_{S}} \mb{\Sigma}_{QD}^{0}(\omega) 
=  - \; \frac{1}{2} \; \left( \begin{array}{cc}  
i\Gamma_{N} & \Gamma_{S} \\ \Gamma_{S} & i\Gamma_{N} 
\end{array} \right)
\label{SAT_limit}
\end{eqnarray} 
therefore the effective quasiparticle energies evolve to $l\omega_{0} 
\pm \sqrt{\tilde{\varepsilon}^{2}+(\Gamma_{S}/2)^{2}}$  and their 
broadening shrinks to $\Gamma_{N}$. 

Focusing on such superconducting atomic limit (\ref{SAT_limit})
we show in Fig.\ \ref{Fig4} the subgap bosonic peaks with 
respect to  $\tilde{\varepsilon}$. In the SET configuration
the energy level $\tilde{\varepsilon}$ would be tunable by 
applying the gate voltage. In particular, these peaks may 
overlap with each other when $\tilde{\varepsilon}\simeq 
\omega_{0}/2$ as reported earlier in the Refs 
\cite{Cho_1999,Sun_2012,Wang_2013}. This effect
can be deduced analytically from 
\begin{eqnarray}
l\omega_{0}+\sqrt{\tilde{\varepsilon}^{2}+(\Gamma_{S}/2)^{2}}
=l' \omega_{0}- \sqrt{\tilde{\varepsilon}^{2}+(\Gamma_{S}/2)^{2}} .
\label{constraint}
\end{eqnarray}
The neighboring peaks ($l'\!=\!l\!+\!1$) overlap when 
$\omega_{0}/2=\sqrt{\tilde{\varepsilon}^{2}+(\Gamma_{S}/2)^{2}}$. 
For small $\Gamma_{S}$ such situation takes place at
$\tilde{\varepsilon}\simeq \frac{1}{2}\omega_{0}$.
Other crossings would be eventually possible for 
the higher-order multiplications of $\omega_{0}/2$.

Figure \ref{Fig4} displays the subgap spectrum $\rho(\omega)$ 
as a function of the coupling $\Gamma_{S}$. From (\ref{constraint})
we conclude that for $\tilde{\varepsilon}=0$ the bosonic peaks 
overlap at $\Gamma_{S}=\omega_{0}$. Energy of these crossing 
points is $\omega=(\frac{1}{2}+l) \omega_{0}$. Here  (for $g=1$) 
we observe four such crossings, but for stronger electron-boson 
couplings a number of the in-gap states and their crossings would increase.

\section{Andreev conductance}

Under nonequilibrium conditions the charge current can be transmitted 
at small voltage $|eV|\!<\!\Delta$ via the Andreev scattering, engaging
the in-gap states. This anomalous transport channel occurs when electrons 
from the metallic lead are converted into the Cooper pairs (propagating 
in superconducting electrode) with the holes reflected back to 
$N$ electrode.  The resulting current $I_{A}(V)$ can be expressed 
by the Landauer-type formula  \cite{transport_formula}
\begin{eqnarray} 
I_{A}(V) = \frac{2e}{h} \int \!\!  d\omega \; T_{A}(\omega)
\left[ f_{FD}(\omega\!-\!eV)\!-\!f_{FD}(\omega\!+\!eV)\right] ,
\label{I_A}
\end{eqnarray} 
with the Fermi-Dirac  function
$f_{FD}(\omega)=\left[ e^{\omega/k_{B}T} + 1 \right]^{-1}$ 
and the Andreev transmittance  \cite{transport_formula}
\begin{eqnarray} 
T_{A}(\omega) = \Gamma_{N}^{2} \;  \left| {\mb F}(\omega) \right|^{2}.
\label{Transmittance_A}
\end{eqnarray} 
Optimal conditions for this subgap transmittance occur when  
$\omega$ coincides with the subgap quasiparticle states. In 
our present case we thus expect a number of such enhancements 
due the bosonic features. Let's remark 
that $T_{A}(-\omega)\!=\!T_{A}(\omega)$ implies the Andreev 
conductance $G_{A}(V) = \frac{ \partial I_{A}(V)}{\partial V}$ 
to be an even function of the bias $V$.

\begin{figure}
\epsfxsize=9.5cm\centerline{\epsffile{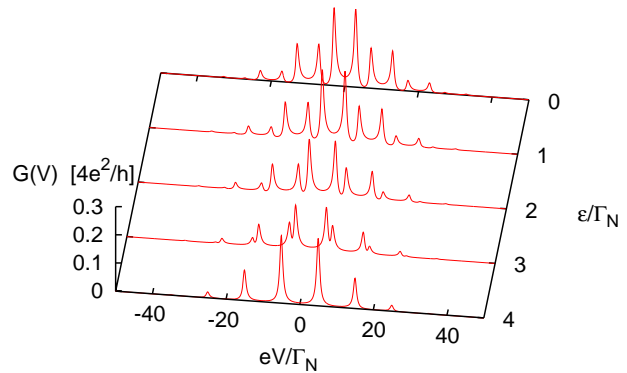}}
\caption{(color online)  The differential Andreev conductance 
$G_{A}(V)$ versus the source-drain voltage $V$ and the QD level 
$\varepsilon$ (tunable by the gate voltage). Results are 
obtained for $T=0$, $g=1$, $\Gamma_S/\Gamma_N$=6, $\omega_0
/\Gamma_N=10$, $\tilde{U}=0$ and $\Delta \gg \Gamma_{S}$. 
Conductance is expressed in units of $4e^{2}/h$.}
\label{Fig5}
\end{figure}

Fig.\ \ref{Fig5} shows the  Andreev conductance as a function
of voltage $V$ applied between the metallic and superconducting 
electrodes. We notice the differential conductance enhancements 
whenever $V$ coincides with the in-gap quasiparticle energies. 
Since $T_{A}(-\omega)\!=\!T_{A}(\omega)$ we observe these maxima 
at $\pm\sqrt{\tilde{\varepsilon}^{2}+(\Gamma_{S}/2)^{2}}\pm 
l\omega_{0}$. They eventually overlap when (\ref{constraint}) 
is satisfied. In particular, for $\Gamma_{S}=6\Gamma_{N}$ 
and $\omega_{0}=10\Gamma_{N}$ the nearest bosonic peaks 
overlap when $\tilde{\varepsilon}=4\Gamma_{N}$. Figure 
\ref{Fig5} clearly shows that the resulting maxima appear 
at $|eV|=\omega_{0}(1/2 + l)$.

\section{Correlation effects}

In various experimental realizations of the quantum dots 
(such as self-assembled InAs islands \cite{Deacon-10}, 
carbon nanotubes \cite{Pillet2013,Schindele2014} or 
semiconducting nanowires \cite{Lee2012,Lee2014}) attached 
to the superconducting leads the energy gap $\Delta$ 
was safely smaller than the repulsion potential $U$. 
For this reason, in the subgap Andreev spectroscopy 
the correlations hardly contributed any Coulomb
blockade. Instead of it, they can eventually induce the  
singlet-doublet quantum phase transition \cite{Bauer-08} 
and/or the Kondo physics \cite{Zitko-15}. In this paper 
we consider the strongly asymmetric coupling $\Gamma_{N} 
\ll \Gamma_{S}$  and focus on the deep subgap regime 
$\Gamma_{N,S} \ll \Delta$, therefore the Kondo-type effects 
\cite{Domanski-EOM,Koerting-10,Rodero-11,Yamada-11,Zitko-15} 
would be rather negligible.

Analysis of such singlet-doublet transition for the vibrating 
quantum dot has been previously addressed  \cite{Zitko_2012}
using the NRG technique. We revisit the same issue here, 
determining the differential Andreev conductance 
(unavailable for the NRG calculations \cite{Zitko_2012}), 
because this quantity could be of interest for experimentalists.  
For the sake of simplicity, we analyze the correlation effects 
in the superconducting atomic limit $\Delta\gg\Gamma_{S}$. 
Hamiltonian of the molecular quantum dot (\ref{afterLF}) 
can be additionally updated with the pairing terms $\frac{1}{2}
\Gamma_{S} \left( \hat{d}^{\dagger}_{\uparrow} 
\hat{d}^{\dagger}_{\downarrow} + \hat{d}_{\uparrow} 
\hat{d}_{\downarrow}\right)$ originating from the static 
off-diagonal parts of the selfenergy matrix (\ref{SAT_limit}). 

\begin{figure}
\epsfxsize=9.5cm\centerline{\epsffile{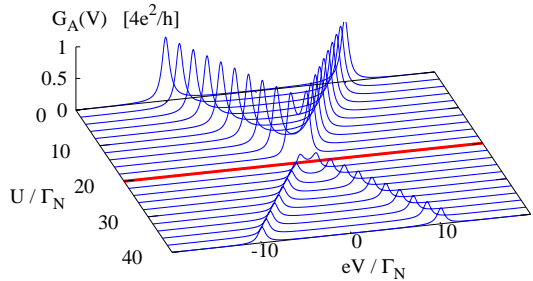}}
\caption{(color online)  The differential Andreev conductance 
$G_{A}(V)$ versus the Coulomb potential $U$ and bias $V$
obtained in the superconducting atomic limit $\Delta \gg \Gamma_{S}$
for $T=0$, $\Gamma_S/\Gamma_N=20$ in absence of the boson mode $g=0$.
The thick (red) line indicates the QPT at $U=\Gamma_{S}$.}
\label{singlet-doublet-free}
\end{figure}

In absence of the boson field (i.e\ for $\lambda\!=\!0$) 
the exact solution of such problem has been discussed by 
a number of authors (e.g.\ see the references cited in 
\cite{Baranski_2013}). The effective quasiparticle energies 
are given by $\pm U/2\pm E_{d}$, where $E_{d}=\sqrt{
(\varepsilon+U/2)^{2}+(\Gamma_{S}/2)^{2}}$. In the 
realistic situations only two branches $\pm \left( 
U/2- E_{d}\right)$ appear in the subgap regime, whereas 
the other high energy states $\pm \left( U/2+ E_{d}\right)$ 
overlap with a continuum  beyond the gap. The quantum phase 
transition (QPT) from the singlet $u\left| 0 \right>
+v\left| \uparrow\downarrow \right>$ to doublet 
$\left| \sigma \right>$ configuration occurs at 
$U/2=E_{d}$ \cite{Bauer-08}. In order to estimate 
quantitatively the Andreev conductance we use 
the off-diagonal Green's function $\mb{f}(\omega)$ 
\cite{Bauer-08,Baranski_2013}, restricting to its 
subgap part
\begin{eqnarray}
\mb{f}_{sub}(\omega) \simeq    
 \frac{\alpha\; uv}{\omega\!+\!\frac{i\Gamma_{N}}{2}
-\left( \frac{U}{2}\!-\!E_{d} \right)} 
- \frac{\alpha\; uv}{\omega\!+\!\frac{i\Gamma_{N}}{2}
+\left(\frac{U}{2}\!-\!E_{d} \right)} 
\nonumber \\
\label{G12_atomic} 
\end{eqnarray}
with the usual BCS coefficient $uv=\Gamma_{S}/4E_{d}$ and 
the spectral weight 
$\alpha = \left[\mbox{\rm exp}\left( {\frac{U}{2k_{B}T}}
\right)+\mbox{\rm exp}\left( \frac{E_{d}}{k_{B}T}\right) 
\right]/{\cal{Z}}$, where ${\cal{Z}}=2\;\mbox{\rm exp}\left( 
{\frac{U}{2k_{B}T}}\right)+\mbox{\rm exp}\left( 
\frac{-E_{d}}{k_{B}T}\right)+\mbox{\rm exp}\left( 
\frac{E_{d}}{k_{B}T}\right)$. The missing part of spectral weight 
$1-\alpha$ belongs to the high-energy states (outside the gap).
At zero temperature this subgap weight changes abruptly from 
$\alpha=1$ (in the singlet state when $U/2<E_{d}$) to 
$\alpha=0.5$ (in the doublet state when $U/2>E_{d}$).

In figure \ref{singlet-doublet-free} we plot the Andreev 
conductance obtained for the half-filled quantum dot
$\varepsilon=-U/2$ (QPT occurs then at $U=\Gamma_{S}$). 
We notice the subgap conductance enhancements around 
$|eV|=U/2-E_{d}$. Yet, exactly at the QPT, both the
singlet and doublet contributions cancel each other. 
Formally, this  is due to the odd (asymmetric) 
structure of the Green's function (\ref{G12_atomic}). 

\begin{figure}
\epsfxsize=9.5cm\centerline{\epsffile{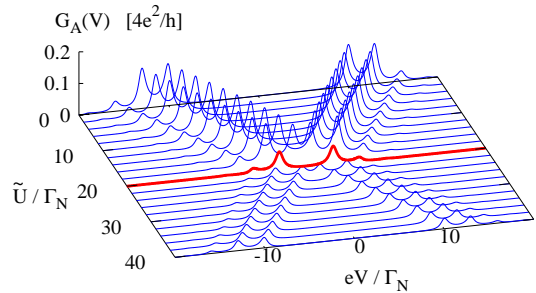}}
\caption{(color online)  The subgap Andreev conductance $G_{A}(V)$ 
as a function of the Coulomb potential $\tilde{U}$ and voltage $V$
obtained for $g=1$, $\omega_0/\Gamma_N=10$ and the same model
parameters as in figure \ref{singlet-doublet-free}.}
\label{singlet-doublet}
\end{figure}

The superconducting atomic limit solution can be generalized 
onto $g\neq 0$ case in a straightforward way. The unitary 
transformation (\ref{Lang_S}) implies $\varepsilon\rightarrow 
\tilde{\varepsilon}$, $U\rightarrow\tilde{U}$ and following
the steps (\ref{G_off}-\ref{SAT_limit}) we can determine 
the off-diagonal Green's function. At zero temperature, 
we find
\begin{eqnarray}
\mb{F}_{sub}(\omega) &\simeq& \alpha\; uv\; \sum_{l=0}^{\infty}
\left\{
 \frac{ e^{-g} \; (-g)^{l} \; /l!}{\omega\!+\!\frac{i\Gamma_{N}}{2}
-\left( \frac{\tilde{U}}{2}\!-\!E_{d} \right) \!+\! s\;l\omega_{0}} 
\right. \nonumber \\ 
&-& \left. \frac{ e^{-g} \; (-g)^{l} \; /l!}{\omega\!+\!\frac{i\Gamma_{N}}{2}
+\left(\frac{\tilde{U}}{2}\!-\!E_{d} \right) \!-\! s\;l\omega_{0}}
\right\} 
\label{G12_atomic_boson}
\end{eqnarray}
with $s\equiv \mbox{\rm sign}(\frac{\tilde{U}}{2}-E_{d})$.

Figure \ref{singlet-doublet} shows the Andreev 
conductance obtained for the half-filled quantum dot using 
$g=1$, $\omega_0/\Gamma_N=10$, $\Gamma_S/\Gamma_N=20$, $T=0$.
The bosonic side-peaks give rise to additional subgap branches, 
similar to what has been reported for the spectral function 
\cite{Zitko_2012}. Right at the QPT, the zero-bias conductance 
again vanishes $G(0)\rightarrow 0$ and we observe only the higher 
order maxima at $|eV|= l \omega_{0}$ (with $l\geq 1$). Away from 
the QPT, the Andreev conductance shows the usual maxima  at  
$|eV|=|\tilde{U}/2-E_{d}|+l\omega_{0}$ whose spectral weights 
depend on $\tilde{U}$ and $l$. 

\section{Summary}

We have investigated the subgap spectrum and transport 
properties of the quantum dot coupled between the metallic and 
superconducting electrodes in presence of the external boson 
mode $\omega_{0}$. We have found that the induced Andreev 
states eventually cross each upon varying the gate potential 
(through $\varepsilon$) or due to the correlations (via 
quantum phase transition from the singlet to doublet 
configurations). We have explored their signatures in 
the measurable charge transport. The tunneling conductance 
of such multilevel 'molecule' shows a series of characteristic 
enhancements, dependent on: the gate voltage with frequency 
$\omega_{0}/2$ (which can be deduced from Eqn.\ \ref{constraint}), 
the bias $V$ applied between external leads (Fig.\ \ref{Fig5}), 
and the correlations (Fig.\ \ref{singlet-doublet}). 
External boson reservoir can thus substantially affect 
the anomalous Andreev current and it can be probed 
experimentally using the low-energy vibrational 
modes or the slowly-varying {\em ac} fields
\cite{Platero_Aguado_2004}.

\section*{Acknowledgment}

This study is partly supported by the National Science Centre (Poland)
under the grant 2014/13/B/ST3/04451. We acknowledge Axel Kobia\l ka
for technical assistence.

\end{document}